\documentclass[aps,prl,showpacs,twocolumn]{revtex4}

\usepackage{graphicx,amsmath,subfigure}

\begin{document}

\title{Observation of a superradiant Mott insulator in the Dicke-Hubbard model}

\author{J. Klinder$^1$, H. Ke{\ss}ler$^1$, M. Reza Bakhtiari$^2$, M. Thorwart$^2$, and A. Hemmerich$^{1,3}$}

\affiliation{
$^1$Institut f\"ur Laser-Physik, Universit\"at Hamburg, Luruper Chaussee 149, 22761 Hamburg, Germany  \\ 
$^2$I.\ Institut f\"ur Theoretische Physik, Universit\"at Hamburg, Jungiusstra{\ss}e 9, 20355 Hamburg, Germany \\
$^3$Wilczek Quantum Center, Zhejiang University of Technology, Hangzhou 310023, China}

\begin{abstract}
It is well known that the bosonic Hubbard model possesses a Mott insulator phase. Likewise, it is known that the Dicke model exhibits a self-organized superradiant phase. By implementing an optical lattice inside of a high finesse optical cavity both models are merged such that an extended Hubbard model with cavity-mediated infinite range interactions arises. In addition to a normal superfluid phase, two superradiant phases are found, one of them coherent and hence superfluid and one incoherent Mott insulating. \end{abstract}

\bibliographystyle{prsty}

\maketitle
The Dicke model, describing the interaction of $N$ two-level atoms with a common mode of the electromagnetic radiation field, is a fundamental paradigm of quantum many-body physics, which despite its long history is still the subject of intensive theoretical research \cite{Dic:54, Dic:64, Hep:73, Gil:78, Bow:79, Gro:82, Dom:02, Ema:03, Gop:09, Nag:10, Stra:11, Bha:12, Bas:12, Pia:13, Tor:13, Kul:13}. As one of its prominent features it exhibits a second order quantum phase transition between a normal phase, in which each atom interacts separately with the radiation mode, and a collective phase in which all atomic dipoles align to form a macroscopic dipole moment \cite{Hep:73, Gro:82}. Only recently, a weekly dissipative variant of this model has been experimentally realized close to zero temperature \cite{Bau:10, Kli:15} by implementing Bose-Einstein condensates inside high finesse optical cavities, which has triggered wide-spread renewed interest \cite{Rit:13}.
   
\begin{figure}
\includegraphics[scale=0.4, angle=0, origin=c]{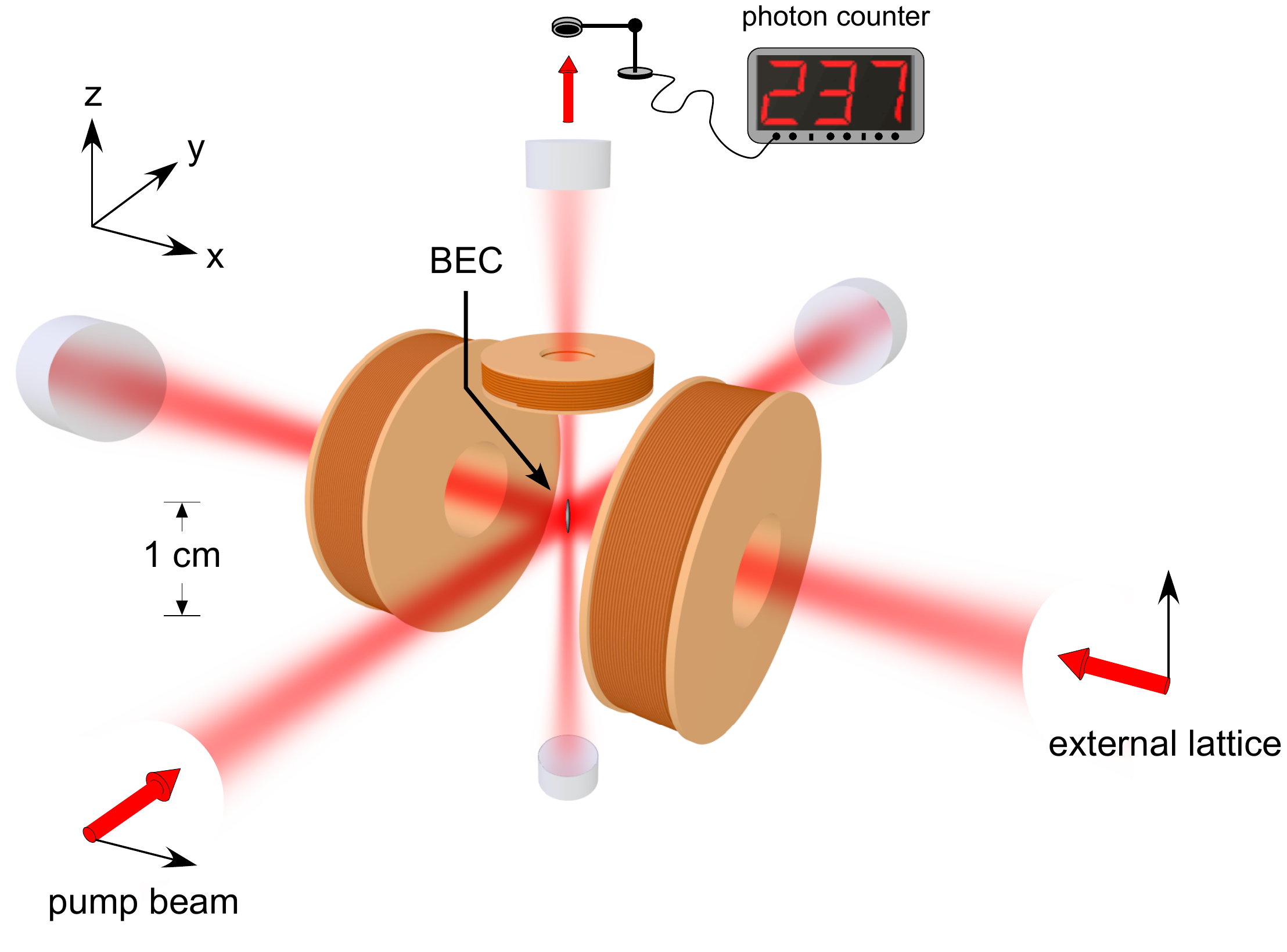}
\caption{Experimental set-up. The beam diameters and the BEC are not drawn to scale.}
\label{fig:Fig.1}
\end{figure}

A similarly elementary model of quantum many-body physics is the Hubbard model, which gives an approximate description of the dynamics of particles on a lattice in terms of the competition of hopping between nearest neighbor sites and on-site collisions \cite{Ger:63, Hub:63}. The Bose-Hubbard model - its bosonic variant - has been originally motivated in the context of superfluid Helium but has received renewed interest after its realization in optical lattices \cite{Jak:98, Gre:02}. At zero temperature, this model is known to possess a quantum phase transition from a superfluid to a Mott insulating ground state \cite{Fis:89} which was confirmed experimentally \cite{Gre:02}. 

In the present work we consider an extended scenario, subsequently referred to as the open Dicke-Hubbard model, which encompasses the physics of both the open Dicke model and the bosonic Hubbard model. Related extensions of Hubbard models have raised wide-spread interest recently due to predictions of highly unconventional phenomena, as for example overlapping, competing Mott-insulator states and strong atom field entanglement \cite{Mas:05, Mas:08, Lar:08, Bha:09}. We study a Bose-Einstein condensate subject to an external lattice potential and interacting with a single light mode of a high finesse optical cavity. In accordance with previous theoretical predictions \cite{Li:13, Bak:15} evidence is found for the existence of three distinct quantum states in the ground state phase diagram: a homogeneous superfluid (HSF) phase, a self-organized superfluid (SSF) phase associated with a spontaneously emerging density grating and a self-organized Mott-insulating (SMI) phase. The phase boundary between the SSF and the SMI phase is observed via a sudden change of phase coherence arising in time-of-flight spectra, which provide an approximate image of momentum space. By successively traversing the phase boundary in both directions we check that coherence is restored as we reenter into the SSF phase. This shows, that the loss of coherence observed in the SMI region cannot be attributed to irreversible heating.

\begin{figure*}[hbt]
\includegraphics[scale=0.48, angle=0, origin=c]{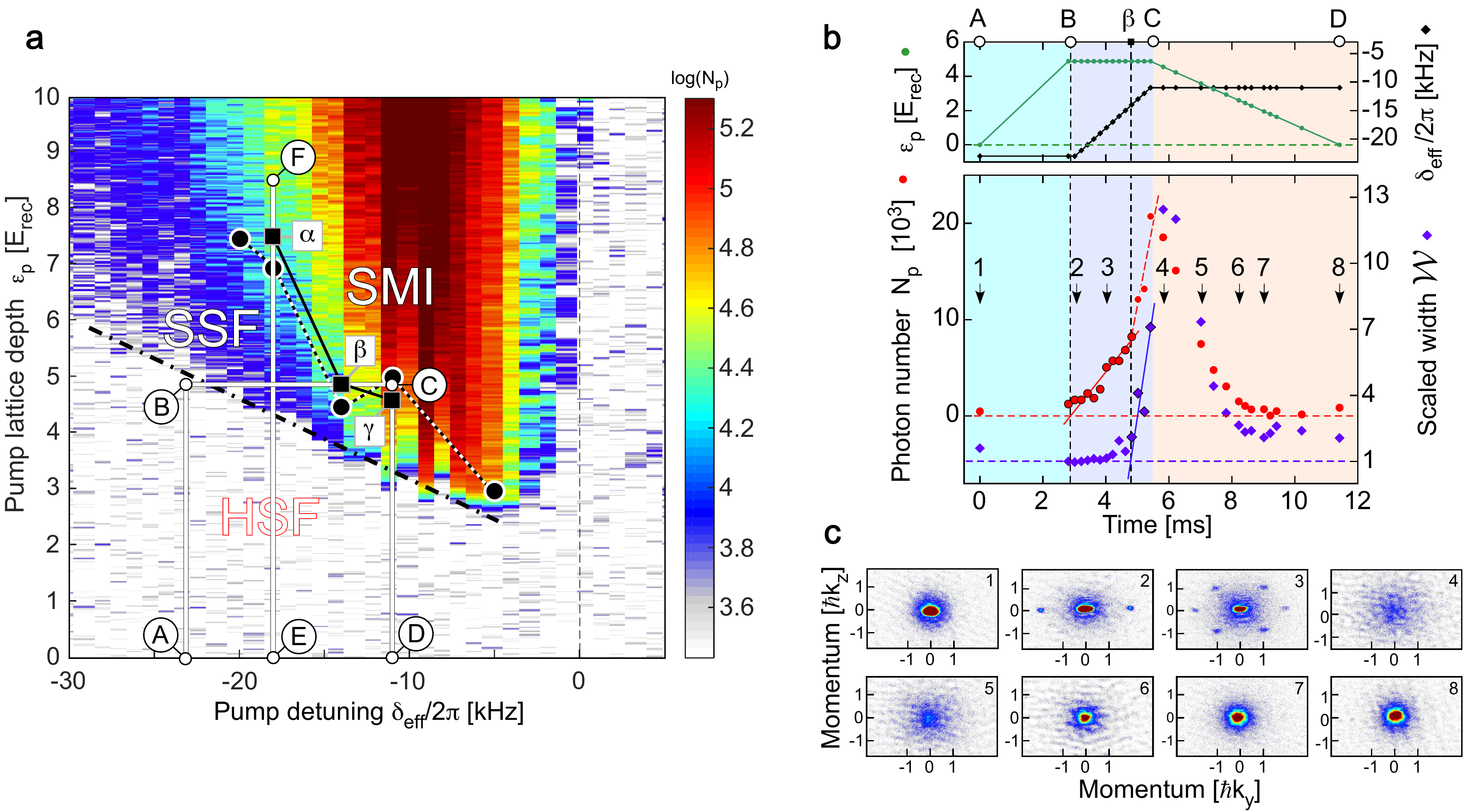}
\caption{(a) The intra-cavity photon number is plotted versus the effective detuning $\delta_{\textrm{eff}}$ and the strength $\varepsilon_p$ of the pump wave. The HSF-SSF phase boundary is highlighted by a thick dashed-dotted black line. Six locations in the phase diagram are highligted by capital letters A, B, C, D, E, F. Three trajectories $\overline{\textrm{ABCD}}$, $\overline{\textrm{DCD}}$, and $\overline{\textrm{EFE}}$ are indicated along which the degree of coherence of the atomic sample is observed. The black squares labeled by greek letters $\alpha, \beta, \gamma$ show the observed boundary between the SSF and the SMI phase. Corresponding BDMFT calculations are shown by the white bordered black disks. (b) The upper panel quantifies the tuning of $\delta_{\textrm{eff}}$ (black squares) and $\varepsilon_p$ (green disks) along the path $\overline{\textrm{ABCD}}$ in (a). In the lower panel the intra-cavity photon number $N_p$ (red disks) and the width (FWHM) $\mathcal{W}$ of the observed zero momentum Bragg resonance (blue diamonds) with regard to the $z$-direction are plotted. The red and blue solid lines are linear fits including the data points highlighted by black margins. The red dashed line corresponds to the white delimited data points. The HSF-SSF and the SSF-SMI phase boundaries are defined by the intersections of the red and blue lines with the $(N_p = 0)$- and $(\mathcal{W} = 1)$-lines, respectively. In (c) momentum spectra (in the ($k_y, k_z$)-plane) are shown recorded at the locations on the path $\overline{\textrm{ABCD}}$ indicated by the black arrows in (b).}
\label{fig:Fig.2}
\end{figure*}

The experimental set-up is sketched in Fig.~\ref{fig:Fig.1}. A cigar-shaped Bose-Einstein condensate (BEC) of $N_a \approx 5 \times 10^4$ $\mathrm{^{87}Rb}$-atoms (prepared in the upper hyperfine component of the ground state $|{F=2,m_F=2}\rangle$) with Thomas-Fermi radii $(3.1, 3.3, 26.8)\,\mu$m is held in a miniaturized magnetic trap with trap frequencies $\Omega_{\mathrm{x,y,z}} / 2\pi = (215.6 \times 202.2 \times 25.2)\,$Hz. The BEC is carefully superimposed to a longitudinal mode ($\approx\, 32\,\mu$m waist) of a high finesse optical cavity extending along the $z$-axis (see Fig.~\ref{fig:Fig.1}). The cavity exhibits a finesse of 344.000, a Purcell factor of 44 \cite{Pur:46} and a field decay rate $\kappa = 2\pi \times 4.45\,$kHz. For a uniform atomic sample and left circularly polarized light, the cavity resonance frequency is dispersively shifted with respect to the case of an empty cavity by an amount $\delta_{-} = \frac{1}{2} N_a \, \Delta_{-}$ with an experimentally determined light shift per photon $\Delta_{-} \approx \,-2\pi \times 0.36\,$Hz. With $N_a = 5 \times 10^4$ atoms $\delta_{-} = -2\pi\times 9$~kHz, which amounts to $-2\,\kappa$, i.e., the cavity operates in the regime of strong cooperative coupling. For $\sigma_{+}$-light $\Delta_{+} \approx \,-2\pi \times 0.16\,$Hz.

The BEC is exposed to two standing wave light fields (see Fig.~\ref{fig:Fig.1}) operating at the wavelength $\lambda = 803\,$nm, i.e., at large detuning to the negative side of the principle fluorescence lines of rubidium at $780\,$nm and $795\,$nm. The pump wave along the $y$-axis may scatter photons into the cavity due to its linear polarization along the $x$-axis. Its tunable strength is parametrized by the spectroscopically determined depth $\varepsilon_p$ of the associated light shift potential in units of the recoil energy $E_\textrm{rec}\equiv \hbar^2 k^2/2m = 2 \pi \hbar \times 3.56\,$kHz associated with photons of wavenumber $k=2\pi/\lambda$. The frequency $\omega_p$ of the pump wave can be precisely tuned relative to the resonance frequency $\omega_c$ of the empty cavity. It is parametrized by the effective detuning $\delta_{\textrm{eff}} \equiv \delta_{c} - \delta_{-}$ with $\delta_{c} \equiv \omega_p-\omega_c$. A second optical standing wave - referred to as the external lattice - is applied along the $x$-axis. It cannot scatter into the cavity due to the linear polarization along the $z$-axis and its fixed frequency detuning of about 360 MHz from the cavity resonance frequency. Thus it merely provides an additional light-shift potential with a fixed well depth of $14\,E_\textrm{rec}$. This lattice acts to split the atomic sample into a collection of effectively two-dimensional sub-samples with any motion along the $x$-direction frozen out. The intra-cavity photon number $N_p$ is precisely determined by counting the photons leaking out through one of the cavity mirrors \cite{Kes:14}.

The experimental scenario discussed here has been previously investigated without an external lattice in the regimes $\hbar \kappa \gg E_\textrm{rec}$ \cite{Bau:10} and $\hbar \kappa \approx E_\textrm{rec}$ \cite{Kli:15}. In both cases the intra-cavity light intensity plotted across the ($\delta_{\textrm{eff}}, \varepsilon_p$)-plane in the negative $\delta_{\textrm{eff}}$ half-plane shows the well known phase transition boundary first predicted by Hepp and Lieb \cite{Hep:73} for a closed system. When $\varepsilon_p$ exceeds a critical value $\varepsilon_{\textrm{p,SSF}}(\delta_{\textrm{eff}})$ the atomic sample scatters photons into the cavity and a self-organized optical lattice emerges. The geometry of this intra-cavity lattice is determined by the interference between the pump field and the intra-cavity field. Via spontaneous symmetry breaking one of two possible lattice potentials is formed trapping the atoms in positions according to the white or black fields of a chequerboard. As is shown in Fig.~\ref{fig:Fig.2}(a), an analogous phenomenon is observed here in presence of the external lattice. The graph shows the observed intra-cavity photon number $N_p$ plotted versus the effective detuning $\delta_{\textrm{eff}}$ and the strength of the pump wave $\varepsilon_p$. It is recorded by repeating the following protocol for varying values of $\delta_{\textrm{eff}}$: 1. a BEC is produced and carefully positioned; 2. the external lattice (along the $x$-direction) is ramped up during 90~ms to $14\,E_{\textrm{rec}}$; 3. the pump strength $\varepsilon_p$ is linearly ramped from zero to $16\,E_\textrm{rec}$ in 10~ms. The observed Hepp-Lieb-Dicke phase boundary $\varepsilon_{\textrm{p,SSF}}(\delta_{\textrm{eff}})$ is highlighted by the black dashed-dotted line.

\begin{figure}
\includegraphics[scale=0.6, angle=0, origin=c]{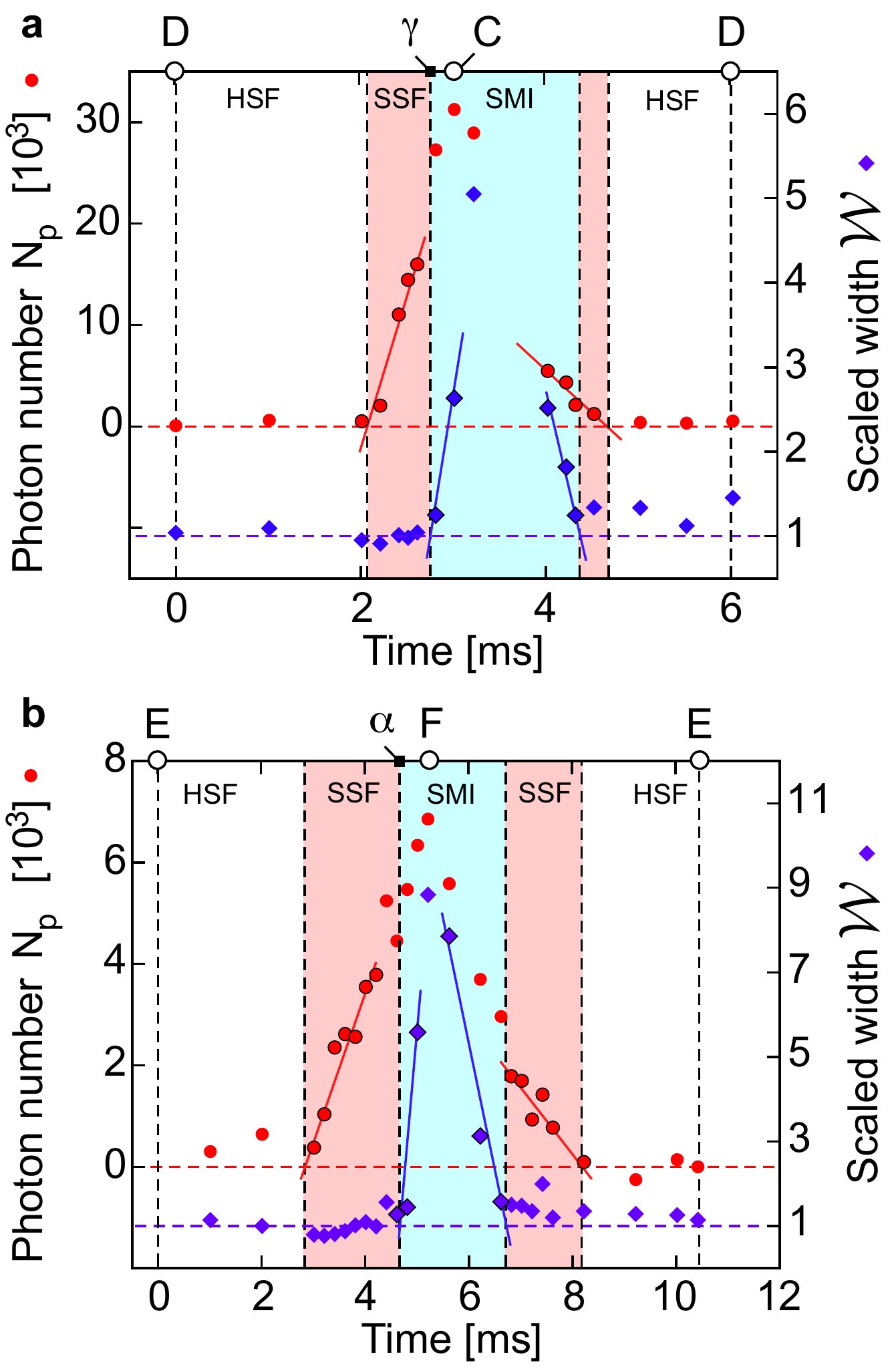}
\caption{Intra-cavity photon number $N_p$ and width (FWHM) of the zero momentum Bragg peak $\mathcal{W}$ plotted versus time for the paths $\overline{\textrm{DCD}}$ (a) and $\overline{\textrm{EFE}}$ (b) indicated in Fig.~\ref{fig:Fig.2}(a). In (a) the pump lattice well depth $\varepsilon_p$ is linearly tuned from zero at D to $5.12\,E_\textrm{rec}$ at C and back to zero at D with constant $\delta_{\textrm{eff}}/2\pi = -11\,$kHz. Correspondingly, in (b) $\delta_{\textrm{eff}}/2\pi = -18\,$kHz and $\varepsilon_p$ is tuned between $\varepsilon_p(E) = 0$ and $\varepsilon_p(F) = 8.45 \, E_{\textrm{rec}}$. The red and blue lines are linear fits including the data points highlighted by black edges. The HSF-SSF and the SSF-SMI phase boundaries are defined by the intersections of the red and blue lines with the $(N_p = 0)$- and $(\mathcal{W} = 1)$-lines, respectively.}
\label{fig:Fig.3}
\end{figure}

Experimentally, we identify a HSF, a SSF and a SMI phase via the following signatures: In the HSF phase no photons are found inside the cavity mode and the momentum spectrum of the atoms shows a pure BEC. In the SSF phase an intra-cavity light field arises, while the momentum spectrum shows sharp Bragg resonances. In the SMI phase the intra-cavity light field increases and the momentum spectrum becomes featureless. By tuning $\delta_{\textrm{eff}}$ and $\varepsilon_p$ we drive the system along the three trajectories $\overline{\textrm{ABCD}}$, $\overline{\textrm{DCD}}$, and $\overline{\textrm{EFE}}$ that intersect the HSF, SSF and SMI phases in different regions of the ($\delta_{\textrm{eff}}$, $\varepsilon_p$)-plane, connecting locations labeled by the capital letters A, B, C, D, E, F in Fig.~\ref{fig:Fig.2}(a). Along these trajectories, in addition to the intra-cavity photon number, we have recorded momentum spectra by rapidly switching off all potentials, allowing for a 25~ms long ballistic expansion and recording an absorption image. Evaluating the widths of the Bragg resonances in these spectra, we obtain information on the degree of phase coherence of the sample along the trajectory. The exemplary case of the path $\overline{\textrm{ABCD}}$, which is traversed in 12~ms, is detailed in Fig.~\ref{fig:Fig.2}(b). The upper panel shows how $\delta_{\textrm{eff}}$ and $\varepsilon_p$ are changed versus time, thus successively passing the points "A,B,C,D". In the lower panel the intra-cavity photon number $N_p$ (red disks) and the width (FWHM) $\mathcal{W}$ of the observed zero momentum Bragg resonance (blue diamonds) with regard to the $z$-direction are plotted. A reference momentum spectrum for $\varepsilon_p=0$ is shown in Fig.~\ref{fig:Fig.2}(c), labeled "1". Note that as the pump strength is raised to $5\,E_{\textrm{rec}}$ without a notable intra-cavity photon number yet arising, $\mathcal{W}$ slightly decreases. This results from the concurrence of the following circumstances: i) The initial condensate without perturbation by the pump wave is strongly elongated in the $z$-direction (aspect ratio $\approx 10$). ii) This elongation is reduced by the effect of the transverse Gaussian trapping potential along the $z$-axis added by the pump wave. iii) The time of flight is too short to be well within the far-field limit with respect to the $z$-axis in the ballistic expansion. We thus have chosen "B" as the reference point for $\mathcal{W}$, setting $\mathcal{W}(B)=1$. Shortly past point "B" the transition into the SSF phase is observed indicated by the sudden increase of the intra-cavity photon number. The width of the zero momentum Bragg peak practically maintains its reference value of unity, which indicates complete coherence, until point "$\beta$" is reached. This is supported by the two exemplary momentum spectra shown for this section in Fig.~\ref{fig:Fig.2}(c), labeled "2" and "3". These spectra show increasing population of higher order Bragg peaks. In "2" the $(\pm 2, 0)\, \hbar k$ Bragg peaks become visible, which result from the $5\,E_{\textrm{rec}}$ deep pump lattice. The intra-cavity contribution to the overall lattice potential is yet negligible. In "3" the visibility of the $(\pm1, \pm1)\, \hbar k$ peaks results from the presence of a notable intra-cavity photon number of $5.6\times10^3$, which yields an overall lattice potential with a well depth of 5.8~$E_{\textrm{rec}}$ with respect to the plane spanned by the cavity and the pump wave. Only when the point "$\beta$" is passed, which corresponds to an intra-cavity photon number $8.2\times10^3$ and 8.0~$E_{\textrm{rec}}$ well depth, a sudden more than ten-fold increase of $\mathcal{W}$ is encountered. At this point a kink in the dependence of $N_p$ upon $\delta_{\textrm{eff}}$ appears, indicating an increase of the superradiant scattering efficiency due to reduced particle number fluctuations \cite{Mek:07}. At point "C" coherence is completely lost as is also directly seen in the momentum spectrum "4" in Fig.~\ref{fig:Fig.2}(c). In the subsequent section $\overline{\textrm{CD}}$ the pump strength is reduced again to zero and $\mathcal{W}$ is observed to decrease again, finally reaching nearly the value initially prepared at point  "A". The significant recovery of coherence indicates that its loss at large lattice depths is not a consequence of excessive heating, but rather indicates the emergence of the SMI state. The particle number $N_a$ at point "D" is reduced to 60 percent of the initial number at point "A". We attribute this to three-body loss associated with the large peak density in the initial BEC of about $1.7 \times 10^{14}\,$cm$^{-3}$, which is compatible with the loss parameter $L = 1.8 \times 10^{-29}\,$cm$^6\,$s$^{-1}$ in Ref.~\cite{Soe:98}. With $N_a = 5 \times 10^4$ and 4 sites per $\lambda^3$ the peak filling factor is 22 particles per lattice site in the trap center. Using a harmonic approximation for the single particle ground state wavefunction for lattice sites near the trap center (at point "$\beta$") leads to the peak density $\rho_{p} = 7.5 \times 10^{15}\,$cm$^{-3}$ and the three-body decay rate $\Gamma_3 \equiv L\, \rho_{p}^2 = 1010\,$s$^{-1}$, i.e., a loss of about 50 percent of the  particles in 10~ms. Without the external lattice the observed particle loss is only on the one percent level \cite{Kli:15}. One may roughly compare the SMI phase with an MI phase in a conventional optical lattice. Assuming a single band Hubbard model, a numerical band calculation yields a nearest neighbor tunneling parameter $J \approx 0.0058\,E_{\textrm{rec}}$ at the point $\beta$. The corresponding collision parameter in harmonic approximation is $U \approx 0.29\,E_{\textrm{rec}}$, such that $U/J \approx 50$. Note, however, that an evaluation accounting for the large filling factors in our experiment should result in a smaller value of $U/J$.

In Fig.~\ref{fig:Fig.3} the two paths $\overline{\textrm{DCD}}$ and $\overline{\textrm{EFE}}$ in the phase diagram in Fig.~\ref{fig:Fig.2}(a) are considered. In these paths the effective detuning is held constant, while the pump lattice depth is linearly ramped from zero to some final value and back to zero. Similarly as in Fig.~\ref{fig:Fig.2}(b), two distinct transition points are observed. Starting from the HSF phase, first the SSF phase is entered indicated by the sudden rise of the intra-cavity photon number, while the coherence measured by $\mathcal{W}$ remains unaffected. Subsequently, a sudden increase of $\mathcal{W}$ shows the emergence of the SMI phase (points "$\gamma$" and "$\alpha$" in Fig.~\ref{fig:Fig.3}(a) and (b)). As the pump strength is reduced again, the coherence is nearly completely restored. In Fig.~\ref{fig:Fig.2}(a) all observed transition points "$\alpha, \beta, \gamma$" for the SSF-SMI transitions are plotted by black squares. 

We have applied bosonic dynamical mean field theory (BDMFT, see \cite{Sno:13} and references therein) to calculate the SSF-SMI transition boundary for 72 particles occupying 18 sites in the SSF phase, adapting previous studies \cite{Li:13, Bak:15} to the configuration of Fig.~\ref{fig:Fig.1}. We assume that the intra-cavity light field  may be integrated out, although the light and matter variables evolve on a comparable time-scale ($\hbar\kappa = 1.25\,E_{\textrm{rec}}$). This greatly facilitates the calculations but cannot account for the dynamical aspects of the experimental system, e.g., the hysteresis observed when the transition boundary between the HSF phase and the SSF phase is crossed in different directions, seen in Fig.~\ref{fig:Fig.3}. Occupations larger than $4$ were not tractable with reasonable calculational expense. The small number of sites and atoms in the calculations were partly compensated by an increased value of the atomic polarizability such that $q \equiv N_a \Delta_{-}/4\kappa$ is close to the experimental value $q = -1$. As the magnitude of $q$ is increased, the calculated SSF-SMI boundary shifts towards lower values of $\varepsilon_p$. The largest value for $|q|$ that we could handle with feasible calculational costs was $|q| = 0.2$. The transition boundary for this case is indicated in Fig.~\ref{fig:Fig.2}(a) by the white-bordered black disks. The BDMFT treatment confirms the presence of a SSF-SMI boundary but does not provide a quantitative description of our observations.

\begin{acknowledgments}
This work was supported by the collaborative research centre DFG-SFB 925 and the Hamburg Centre for Ultrafast Imaging (CUI). We acknowledge useful discussions with W. V. Liu, C. Morais Smith, L. Mathey, and C. Zimmermann.
\end{acknowledgments}


\begin{thebibliography}{99}
\bibitem{Kes:14} 
The following calibration procedure is used: a unidirectional pump pulse is applied (along the $y$-axis) such that all atoms are transferred from the condensate mode $(p_y, p_z) = (0,0)$ into the $(p_y, p_z) = (2 \hbar k, 0)$-momentum class by scattering two photons via intermediate population of the $(\hbar k, \pm \hbar k)$-classes (cf. Ref.~\cite{Kes:14b}). With $N_a$ atoms in the cavity, we then know that $2 N_a$ photons have been scattered into the cavity.
\end{thebibliography}
\end{document}